\documentclass[twocolumn]{aastex62}
\usepackage{latexsym}
\usepackage{graphicx}
\usepackage{color}
\usepackage{amsmath}

\usepackage{hyperref}
\usepackage{bm}
\usepackage{acronym}
\usepackage{url}
\usepackage{textcomp}

\begin{document}

\reportnum{arXiv:1806.02822}
\reportnum{LIGO-P1800125}
\reportnum{INT-PUB-18-031}

\title{
Evidence for a minimum ellipticity in millisecond pulsars}

\author{G.~Woan}
\author{M.~D.~Pitkin}
\affiliation{SUPA, School of Physics \& Astronomy,
University of Glasgow, Glasgow G12 8QQ, United Kingdom}
\email{graham.woan@glasgow.ac.uk}

\author{B.~Haskell}
\affiliation{Nicolaus Copernicus Astronomical Center of the Polish Academy of
Sciences, Ulica Bartycka 18, 00-716 Warszawa, Poland}

\author{D.~I.~Jones}
\affiliation{Mathematical Sciences and STAG Research Centre,
University of Southampton, Southampton SO17 1BJ, United Kingdom}

\author{P.~D.~Lasky}
\affiliation{School of Physics and Astronomy, Monash University,
Vic 3800, Australia}
\affiliation{OzGrav: The ARC Centre of Excellence for Gravitational
Wave Discovery, Monash University, Vic 3800, Australia}

\date{\today}

\begin{abstract}
Neutron stars spin down over time due to a number of energy-loss processes.
We provide tantalizing population-based evidence that millisecond pulsars (MSPs)
have a minimum ellipticity of $\epsilon\approx10^{-9}$  around their spin
axis and that, consequently, some spin down mostly through gravitational-wave
emission. We discuss the implications of such a  minimum ellipticity in terms
of the internal magnetic field strengths and nuclear matter composition of
neutron stars and show it would result in the Advanced LIGO and Virgo
gravitational-wave detectors, or their upgrades, detecting gravitational
waves from some known MSPs in the near future.
\end{abstract}


\section{Introduction}
Any rotating system generates gravitational waves if its mass is not arranged
symmetrically around the axis of rotation, and the recent discovery of
gravitational waves from the coalescing binary neutron star system GW170817
is a clear example of this general result \citep{2017PhRvL.119p1101A}. Even
isolated, but rapidly-rotating, neutron stars (if sufficiently asymmetric)
could generate gravitational-wave signals detectable on Earth. Indeed,
approximately 451 of the 2\,636 known radio and X-ray pulsars would generate
continuous quadrupolar ($l=2$) emission that falls within the frequency range
of ground-based gravitational-wave detectors, prompting deep searches for
these signals in data from both LIGO and Virgo
\citep[e.g.,][]{2014ApJ...785..119A, 2017ApJ...839...12A}.\footnote{These
numbers assume all pulsars with rotational frequencies greater than 10\,Hz
are within the sensitive frequency range of gravitational-wave detectors.
They are taken from version 1.58 of the Australia Telescope National Facility
(ATNF) Pulsar Catalog
\citep{2005AJ....129.1993M}.} To date no signals have been detected, but
these investigations have placed stringent upper limits on the mass
quadrupole of many known pulsars, and in several instances limits on the
fraction of overall luminosity that can be attributed to gravitational-wave
emission \citep{2017ApJ...839...12A}.

The strength of the gravitational-wave emission from a neutron star depends
on both the degree of asymmetry and the rotation rate. For a non-precessing
triaxial star, rotating about its $z$ principal axis, the asymmetry is
dominated by the $m=2$ mass quadrupole $Q_{22}$ and characterised by the
moment of inertia ellipticity $\epsilon$ \citep[see,
e.g.,][]{2005PhRvL..95u1101O}
\begin{equation}
\epsilon = \frac{I_{xx}-I_{yy}}{I_{zz}} = \frac{Q_{22}}{I} \sqrt{\frac{8\pi}{15}},
\end{equation}
where $I_{ii}$ are the principal moments of inertia and $I_{xx}\simeq
I_{yy}\simeq I_{zz} = I$. A star of rotational period $P$ and ellipticity
$\epsilon$ will generate gravitational waves of period $P/2$  and with a
gravitational luminosity of
\begin{equation}
L_\text{GW} =  \frac{2048\pi^6}{5} \frac{G}{c^5} I^2\epsilon^2 P^{-6}
   \simeq 10^{29}\left(\frac{\epsilon}{10^{-9}}\right)^2
   \left(\frac{1\,\text{ms}}{P}\right)^{6}\!\! \text{W}.
\end{equation}
The luminosity is proportional to the square of the third time derivative of
the (reduced) moment of inertia tensor, giving it a strong period dependence.
This loss of energy acts as a rotational brake on the neutron star and, for a
pulsar, contributes to an observed rate of change of period, or ``spin-down''
rate.

However, gravitational radiation is usually expected to be a relatively small
contribution to the overall spin-down. Pulsars are thought to have strong
external magnetic fields, with spin-down rates dominated by magnetic dipole,
rather than gravitational quadrupole, radiation and with additional braking
mechanisms present including wind-induced mass-loss.

A pulsar spinning at an angular frequency $\omega$ has a braking index $n$
that satisfies $\dot\omega \propto -\omega^n$ (or $\dot{P} \propto
P^{-(n-2)}$), and $n$ can be measured for some pulsars, revealing a range of
values (see, e.g., Table~1 of \citealt{2015MNRAS.446..857L} and references
therein).

The process of spin-down is clearly complicated, but an isolated,
magnetically braked, rigid rotator would have a braking index of three. If
gravitational emission is the dominant process we have a ``gravitar'' with a
braking index of five \citep{2005MNRAS.359.1150P}.

Right from the start of pulsar studies, it was noted by \citet{Ferrari69}
that the overall spin-down in even the simplest systems would have
contributions from both, so that at the very least
\begin{equation}
\dot \omega = - \alpha \omega^3 - \beta \omega^5.
\label{overall_spindown}
\end{equation}
LIGO and Virgo observations have shown the gravitational contribution to the
spin-down of the Crab pulsar to be tiny \citep{2017ApJ...839...12A}, but its
spin-down rate is high and it would need a possibly unphysical $\epsilon\sim
10^{-4}$ to be dominated by gravitational-wave emission. Conversely,
gravitational observations have constrained the ellipticity of some
millisecond pulsars (MSPs) to $\epsilon \sim 10^{-7}$ or less
\citep{2017ApJ...839...12A}. Such ellipticities are well within the bounds
set by likely neutron star equations of state
\citep{2005PhRvL..95u1101O,2013PhRvD..88d4004J}, and the short rotational
period of MSPs would make them relatively luminous gravitational wave
sources.

The spin periods ($P$) and spin-down rates ``Pdot'' ($\dot P$) of the pulsar
population are traditionally displayed on a P-Pdot diagram \citep[e.g.,
Figure~1.13 of][]{2004hpa..book.....L} showing how known pulsars cluster in
particular regions of this parameter space, and new surveys have bolstered
the number of sources considerably.  The clustering is thought to result from
a mixture of observational selection effects and underlying physics.

In this Letter, we discuss whether there is evidence for a new cutoff
emerging in the diagram, at short-period and low-spin-down rate, caused by
gravitational-wave emission and consistent with a \emph{minimum} ellipticity
for MSPs. Such a cutoff would correspond to a population of rapidly-rotating
gravitars, sufficiently luminous to be detectable by current or future
ground-based gravitational observatories.

\section{Data and model}\label{data_and_model}
To investigate this apparent cutoff we will consider known pulsars with
periods $P<10$\,ms and period derivatives $\dot P<10^{-18}$\,s\,s$^{-1}$. The current
ATNF pulsar catalogue \citep{2005AJ....129.1993M} contains 199 pulsars that
fulfil these criteria. The observed values of their period derivatives will,
to varying degrees, be contaminated by radial accelerations due to proper
motion (the Shklovskii effect), differential Galactic rotation
\citep{1991ApJ...366..501D} and, for pulsars in globular clusters, local
forces \citep{Freire17}.   We therefore need to consider these effects and,
as far as possible, work with the true (i.e., \emph{intrinsic}) period
derivatives.

\begin{figure}
\includegraphics[width=\columnwidth]{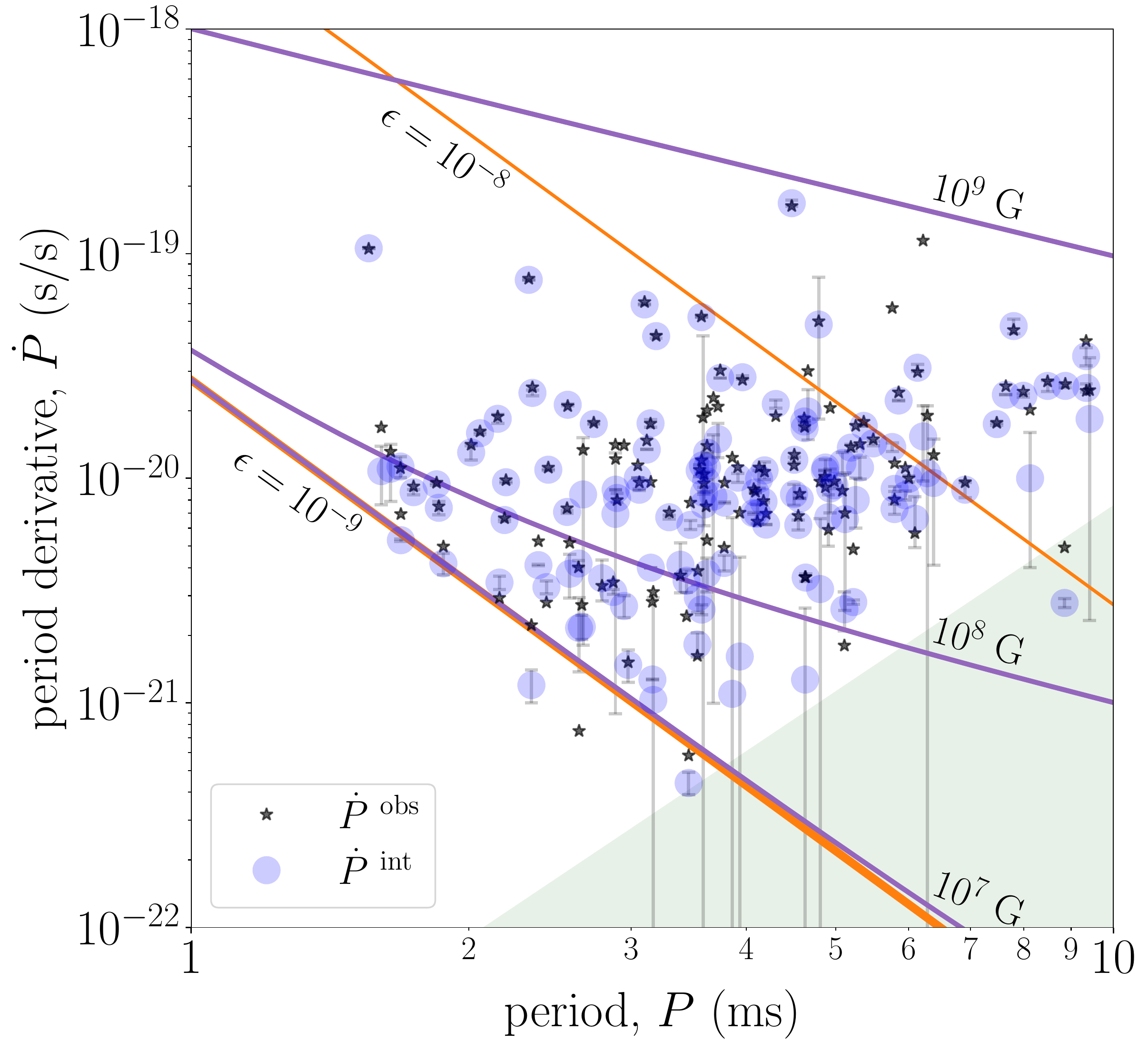}
\caption{\label{fig2}
Observed (black stars) and intrinsic (blue circles) period derivatives of
the MSPs in our sample vs.\ period (i.e., the bottom-left corner of the
standard P-Pdot diagram) excluding pulsars in globular clusters. The intrinsic
period derivatives include corrections for differential Galactic
rotation and Shklovskii effects. Straight (orange) lines show where canonical
gravitars with ellipticities of $10^{-8}$ and  $10^{-9}$ would sit.  Note the
sudden fall in source numbers below $\epsilon\simeq 10^{-9}$. Also shown are
curves (blue) of constant surface magnetic field $B_\text{s}$, assuming
spin-down is in accordance with Equation~(\ref{BGspindown}) and that all the
pulsars have a common ellipticity of $10^{-9}$.  The shaded area corresponds
to the death line exclusion region of \citet{2000ApJ...531L.135Z}. }
\end{figure}

We begin by excluding all 59 globular cluster pulsars from our sample. We
also exclude PSR J1801\textminus3210, which shows no measurable period
derivative but which is thought to be affected by Galactic acceleration
\citep{Ng14}, and J1400\textminus1431, for which there is only an upper limit
on Pdot \citep{2017ApJ...847...25S}; this leaves 128 MSPs. For 28 of these we use
intrinsic period derivatives from the literature, already corrected for
Shklovskii and differential Galactic rotation effects using parallax-based
distance estimates (and in many cases also corrected for Lutz-Kelker bias)
\citep{2012ApJ...756L..25D, Ng14, 2016MNRAS.458.3341D, 2018MNRAS.475..469S}.
For the remaining pulsars we calculate the corrections using the model of
Damour and Taylor but with a Galactic radius of 8.3\,kpc. We use
parallax-derived distances when available, either from the literature
(corrected for Lutz-Kelker bias; \citealp{2016MNRAS.458.3341D,
2016MNRAS.455.1751R}) or using the values given by the ATNF pulsar catalog
\citep{2005AJ....129.1993M}. If no parallax distance is known we use the
best-estimate distance given in the ATNF catalogue, which by default uses the
measured dispersion measure and the Galactic electron density model of
\citet{2017ApJ...835...29Y}. The P-Pdot diagrams using intrinsic (circles)
and observed (stars) period derivatives are shown in Figure~\ref{fig2}.

We can cast Equation~(\ref{overall_spindown}) into a standard form by assuming
a neutron star of radius $R$, with magnetic spin-down due to vacuum dipole
radiation and surface magnetic field intensity $B_\text{s}$:
\begin{equation}\label{eq:spindown}
\dot P = \frac{32\pi^3 R^6}{3Ic^3\mu_0} B_\text{s}^2 P^{-1} +
         \frac{512\pi^4 G I}{5c^5} \epsilon^2 P^{-3}.
\end{equation}
Using canonical values for the radius (10\,km) and moment of inertia
($10^{38}$\,kg\,m$^2$) of the neutron star this becomes
\begin{equation}
\begin{split}
  \left(\frac{\dot P}{10^{-20}\,\text{s/s}}\right)
          = & 0.98\left(\frac{1\,\text{ms}}{P}\right)
          \left(\frac{B_\text{s}}{10^8\,\text{Gauss}}\right)^2 \\
            & + 2.7 \left(\frac{1\,\text{ms}}{P}\right)^3
                               \left(\frac{\epsilon}{10^{-9}}\right)^2.
\label{BGspindown}
\end{split}
\end{equation}
Canonical gravitars (i.e., neutron stars obeying Equation~(\ref{BGspindown})
but with negligible magnetic field) would fall on the straight orange lines
in Figure~\ref{fig2}. The blue curves show where canonical pulsars with
$\epsilon=10^{-9}$ would be located for different values of surface magnetic
field.  Pulsars with a range of ellipticities and magnetic fields would also
fall in this region, consistent with Equation~(\ref{BGspindown}).  The number
of pulsars involved is small, but there is some evidence for a cutoff in the
population below the gravitar $\epsilon=10^{-9}$ line in the limit of low
magnetic field, consistent with the notion that MSPs have a residual
ellipticity that does not tend to fall below this level. Although such
weakly magnetized pulsars would constantly radiate the equivalent of several
solar-luminosities in gravitational waves, their reservoir of rotational
kinetic energy is huge ($\sim 10^{43}$\,J) and their gravitational spin-down
age, $P/(4\dot P)$, is $10^8$--$10^{10}$\,years.  These limiting neutron stars
must still have sufficient magnetic field to be seen as pulsars, but their
dominant spin-down mechanism would be gravitational. The two pulsars below
the $\epsilon=10^{-9}$ line are PSR~J2322\textminus2650
\citep{2018MNRAS.475..469S}, a recently discovered MSP with low radio
luminosity and a low-density planetary-mass companion, and
PSR~J1017\textminus7156 \citep{Ng14}.

\subsection{Fitting the P-Pdot distribution to the model}\label{sec:fitting}
The model described by Equation~(\ref{BGspindown}) assumes the same moment of
inertia for all neutron stars. In reality, the moment of inertia of a star
depends on its mass via the equation of state, and particularly massive or
light neutron stars may have larger (or smaller respectively) values of Pdot
than those predicted by Equation~(\ref{BGspindown}) of up to a factor $\approx
2$ \citep{2008ApJ...685..390W}, leading to additional scatter in the P-Pdot
diagram. The significance of the apparent cutoff therefore needs to be
considered more carefully, with these effects in mind.

We therefore construct a simple model of the distribution of pulsars over the
P-Pdot plane. We assume a priori that spin-down rates are distributed
uniformly in log-space,  with a  lower cutoff that follows a power law of
the form $\dot{P} = kP^{-(n-2)}$, corresponding to a braking index of $n$. A
special case of this cutoff process would be gravitational radiation
(Equation~(\ref{eq:spindown}) with $B_{\rm s}=0$) from a common ellipticity.
The lower-right corner of the P-Pdot plane is also largely free of
sources and is delimited by the ``death line,'' below which neutron stars are
not observable as pulsars. We use the death line defined in Equation~(3) of
\citet{2000ApJ...531L.135Z} to exclude the region below $\dot{P} = L_\text{d}
(P/1\,\text{s})^{11/4}$\,s/s, where $L_\text{d} = 10^{-14.62}$, and find the
common cutoff process that best explains the observed distribution. There
are also few MSPs with $\dot P \gtrapprox 10^{-19}$\,s\,s$^{-1}$, which will depress
the model evidence values but not significantly affect comparisons between
parameterizations of the lower cutoff.

We assign a Gaussian likelihood for the true period derivative of the $j$th
pulsar $\dot P_j$, given our measurement of its intrinsic Pdot and its
uncertainty, so that
\begin{equation}
    p(d_j|\dot{P}_j)  \propto
\exp\left[-\frac{1}{2\sigma_j^2}\left(\dot{P}_j - \dot P^\text{int}_j\right)^2\right],
\end{equation}
where the values of $d_j \equiv \{\dot P^\text{int}_j, \sigma_j\}$ are set by
the intrinsic Pdot estimation procedure described in
Section~\ref{data_and_model}, and shown in Figure~\ref{fig2}. If
uncertainties for the intrinsic Pdot are available from the literature, then we use
those (see Section~\ref{data_and_model}). Otherwise we combine, in
quadrature, the measurement uncertainties on Pdot, the Shklovskii correction,
and the Galactic correction, and these comprise the error bars in
Figure~\ref{fig2}. We assign a conservative 50\% uncertainty on any distances
derived from dispersion measure. We incorporate the cutoff process and death
line into the log-uniform prior on $\dot{P}_j$ as a common threshold
$\dot{P}_j^\text{th}(P_j)$, such that
\begin{equation}
p(\dot{P}_j|\dot{P}_j^{\rm th}) = \begin{cases}
\left[\dot{P}_j\ln\frac{{\dot{P}^{\rm max}}}{\dot{P}^\text{th}_j}\right]^{-1} &
      \text{if } \dot{P}_j^{\rm th} \leqslant \dot{P}_j \leqslant \dot{P}^{\rm max}, \\
    0 & \text{otherwise},
    \end{cases}
\end{equation}
where  $\dot{P}^\text{max}$ is the maximum value of $\dot{P}^\text{int}$ in
our sample.

The exclusion region delimited by $\dot{P}^\text{th}(P)$ depends on both our
new cutoff line and the death line, and we take $\dot{P}^{\rm th}(P)$ to be
the maximum of $\dot{P}(P)$ defined by the cutoff line (parameterized by $k$
and $n$) and the death line and incorporate the moment of inertia of each
pulsar into the cutoff process by setting $k_j = I_j k'$. We can then set a
Gaussian prior on $I_j$, with a common mean $\mu_{I}$ and standard deviation
$\sigma_{I}$ over all pulsars and choose $\mu_{I} = 2\!\times\!10^{-38}\,{\rm
kg}\,{\rm m}^2$ and $\sigma_{I} = 3\!\times\!10^{37}\,{\rm kg}\,{\rm m}^2$
using the ranges shown in Figures~4~\&~7 of \citet{2008ApJ...685..390W}, with
hard bounds of $[1\!\times\!10^{38}, 3\!\times\!10^{38}]\,{\rm kg}\,{\rm
m}^2$. We also incorporate uncertainty in the position of the death line by
allowing the coefficient $L_\text{d}$ to be a free parameter with a prior
that is uniform in log-space between $[10^{-15.62}, 10^{-14.62}]$, which
spans most of the range of death lines in \citet{2000ApJ...531L.135Z}.
Therefore, for a given pulsar, $\dot{P}_j^\text{th}$ is defined by $P_j$,
$k'$, $I_j$, $n$ and $L_\text{d}$.

For each pulsar, we first marginalize over $\dot{P}_j$ and $I_j$ to give
\begin{equation}\label{eq:pdotlikejoint}
 p(d_i|L_\text{d}, k', n) =
    \iint
     p(d_i|\dot{P}_j) p(\dot{P}_j|\dot{P}^\text{th}_j)
     p(I_j) \,\text{d}\dot{P}_j\,\text{d} I_j.
\end{equation}
We then form a joint likelihood over all $N$ pulsars in our sample and
marginalize over $L_\text{d}$ and $k'$ to give the marginal likelihood, or
evidence, for the model as a function of the cutoff process braking
index $n$, giving
\begin{equation}
    p(\boldsymbol{d}|n) = \iint \prod_{j=1}^N p(d_j|L_\text{d}, k', n)
    p(L_\text{d}) p(k') \,\text{d}L_\text{d}\,\text{d}k',
\end{equation}
where $\boldsymbol{d}$ is the combined data $\{d_i\}$, and the prior on $k'$,
$p(k')$, is log-uniform over a range for which
$\ln{(k'_\text{max}/k'_\text{min})} \approx 44$.

Using all the data from pulsars for which we have confident measurements of
intrinsic Pdot (shown in Figure~\ref{fig2}) we can therefore determine Bayes
factors,
\begin{equation}
{\cal B}(n) = \frac{p(\boldsymbol{d}|n)}{p(\boldsymbol{d}|k'=0)},
\end{equation}
comparing models with and without a common cutoff process other than the death
line. If we take the cutoff process to be due to an $n=5$ process, then we find that
$\cal B$ is hugely in favour of a distribution with a non-zero ellipticity (a
factor of $\sim 6\,400$), corresponding to an ellipticity of $\sim
5.3^{+0.4}_{-0.7}\!\times\!10^{-10}$. We note here that as we have taken a
distribution for $I$ centred at twice the canonical value, this ellipticity
corresponds to the same mass quadrupole, and therefore gravitational wave
amplitude, as choosing an ellipticity of $\sim 10^{-9}$ and using the canonical
moment of inertia.

Comparing the evidence for an $n=5$ process to an $n=3$ process we find that the
former is favoured by a factor $\sim 35$. In fact, we find that a process with $n
\approx 5.6_{-0.4}^{+1.1}$ is the most favoured case, being $\sim 10,300$
times more likely than having no cutoff, and $\sim 55$ times more likely than
an $n=3$ process. We conclude, therefore, that the observed fall-off in the number
of sources at the bottom-left of Figure~\ref{fig2} is highly significant and
much more likely due to a common ellipticity amongst MSPs than due to a common
minimum surface magnetic field.

Although this simple model explains the short-period/low Pdot cutoff nicely,
there could be other processes and selection effects involved that we are not
modeling. Very low intrinsic period derivatives are difficult to measure,
but the most important corrections (applied above) are thought to be
well-understood. As is clear in the case of globular cluster pulsars,
significant but uncorrected additive errors in Pdot tend to scatter pulsars
well away from the small region of (linear) parameter space containing the
cutoff boundary, and such pulsars would not influence the analysis. It should
also be noted that a relatively long time baseline is required to measure low
Pdot values, but this becomes somewhat easier for shorter period pulsars, so
these should not be preferentially excluded by this requirement.

Although our Bayes factors are naturally modulated by prior assumptions, they
strongly support the apparent power-law cutoff one sees in the P-Pdot plot,
and show it to be consistent with a limiting braking index of $n=5$, rather
than $n=3$. Evidence for or against a simple cutoff would be strengthened by
a larger sample of pulsars with periods below $\sim 3$\,ms, where the death
line has less of an influence. Similarly, a larger number of pulsars close to
the cutoff line might indicate a form more complex than a power law.
However, with the current limited sample size, more complex models are
naturally disfavored by the Occam factor.

\section{Discussion}
It is reasonable to ask what could cause the minimal ellipticity  described
in Section~\ref{data_and_model}.  Magnetic field evolution in MSPs is not
well-understood; however, these stars have relatively small external dipole
magnetic fields ($B_\text{s}\sim10^{8}$\,G) compared to the younger
population of radio pulsars ($10^{11}\lesssim B_\text{s}/{\rm
G}\lesssim10^{13}$).

One possibility is that the external magnetic field becomes buried while the
system is undergoing accretion from the binary companion. If
this is the case, one may expect an internal magnetic field on the order of
$10^{11}$\,G \citep{VM09}. MSPs are old, cold
stars, and the protons in their cores are expected to form a type II
superconductor. For such stars, the ellipticity is linear in the internal
magnetic field strength. The exact value is model-dependent, but generally of
the order of~\citep{2012MNRAS.419..732L, 2014MNRAS.437..424L}
\begin{equation}
\epsilon\sim10^{-8}\left(\frac{\langle B_\text{i} \rangle}{10^{12}\,\text{G}}\right)
    \left(\frac{\langle H_\text{c}\rangle}{10^{15}\,\text{G}}\right),
\end{equation}
where $H_\text{c}$ is the lower critical field for superconductivity. An
ellipticity of $10^{-9}$ would therefore be consistent with a buried field of
$B \sim10^{11}$\,G, itself consistent with the field strengths observed in
the general pulsar population.

Another possible explanation for such a minimum ellipticity could be strain
in the elastic crust; a value of $\epsilon\approx 10^{-9}$ can be
accommodated with dimensionless strain levels of around $10^{-4}$
\citep{2000MNRAS.319..902U}, comfortably below the breaking strains indicated
by molecular simulations \citep{2009PhRvL.102s1102H}.  Such strains might be
produced by asymmetries in the accretion process through which these MSPs
were originally spun-up \citep{1998ApJ...501L..89B, 2000MNRAS.319..902U}, or
else by asymmetric crustal fracture during the accretion spin-up
\citep{Fattoyev18} or the subsequent post-accretion spin-down
\citep{1971AnPhy..66..816B}.

At this point it is interesting to ask whether there is any evidence for such a
minimum ellipticity in the population of accreting millisecond X-ray pulsars in
low-mass X-ray binaries, which are thought to be the progenitors of millisecond
radio pulsars. They have also long been studied as a source of gravitational
waves because their spin periods appear to cluster around $P\approx 2$\,ms ---
well below the theoretical Keplerian breakup frequency to which a neutron star
could be spun-up \citep{2017ApJ...850..106P}. Gravitational waves could be
providing an additional spin-down torque that balances the spin-up torque due
to accretion. This leads to equilibrium at the observed periods for
ellipticities of $\epsilon = 10^{-8}$ in most of the population
\citep{1998ApJ...501L..89B,2000MNRAS.319..902U}. There are however, two systems
close to the minimum observed period that are transitional pulsars; i.e., are
making the transition from being accretion powered X-ray pulsars to being
rotationally powered millisecond radio pulsars:  J1023+0038 and J1227\textminus4853. Both
are accreting at low rates, so that the accretion torque would be balanced for
$\epsilon\approx 2\times 10^{-10}$ \citep{2017ApJ...850..106P}. Nevertheless,
these systems cannot be in spin-equilibrium as they are observed to spin-down in
radio, and J1023+0038 not only appears in our sample close to the cutoff, but
it is also observed to spin-down $\approx 30\%$ faster in X-ray
\citep{2016ApJ...830..122J}. Intriguingly, this effect could be explained by
gravitational-wave emission \citep{2017PhRvL.119p1103H}, suggesting that these
systems may be the progenitors of the pulsars close to the cutoff in the P-Pdot
diagram.

\section{Implications for future gravitational-wave detections}
We now determine the corresponding gravitational-wave signal strength from our
selected MSPs \citep[using, e.g., Equation~(3) of][]{2014ApJ...785..119A} under
two models: (i)~that they each have an ellipticity of $10^{-9}$ as considered
above and (ii) that they are all maximal gravitars, with signal strengths set by
their spin-down limits. In the first case we are able to re-include those
pulsars in globular clusters, as knowledge of their true period derivatives is
not required. We use the canonical moment of inertia and distances obtained as
described in Section~\ref{data_and_model}, keeping in mind that there could be
factor-two uncertainties in both these values. In the gravitar case we calculate
spin-down limits only for those pulsars not associated with globular clusters
and for which we have confident measurements of their intrinsic Pdot, as
discussed in Section~\ref{data_and_model} and shown in Figure~\ref{fig2}.

Under these models we have estimated the signal-to-noise ratios \citep[see
Equation~(2) of][]{2011MNRAS.415.1849P} that pulsars with rotation periods less
than 10\,ms  would have in future networks of gravitational-wave detectors,
assuming fully-coherent targeted searches. Their signal-to-noise ratios depend
on the (unknown) inclination of each pulsar's rotation axis with respect to the
line-of-sight.  We have calculated signal-to-noise ratios for the angle-averaged
case, a factor of $\sim 1.69$ times below the best-case, assuming a uniform
prior on orientation \citep{2011MNRAS.415.1849P}. The networks we have
considered are: (a) two advanced LIGO (aLIGO) detectors, in Hanford and
Livingston, operating at design sensitivity
\citep{aLIGOdesign,2015CQGra..32g4001L} with the advanced Virgo (AdV) detector
also operating at design sensitivity \citep{AdVdesign,2015CQGra..32b4001A}; (b)
this configuration combined with an equivalent LIGO detector in India
\citep{2013IJMPD..2241010U} and the KAGRA detector \citep{2013PhRvD..88d3007A}
at design sensitivity; (c) two upgraded aLIGO+ \citep{ISWP:2016} detectors,
together with AdV; (d) Cosmic Explorer \citep{2017CQGra..34d4001A}; (e) the
Einstein Telescope in its ET-D configuration \citep{2011CQGra..28i4013H,
2012CQGra..29l4013S}, assuming three co-located detectors. In all cases we use a
one-year observation period and coherently combine data from all detectors in a
network.

\begin{figure}
    \includegraphics[width=\columnwidth]{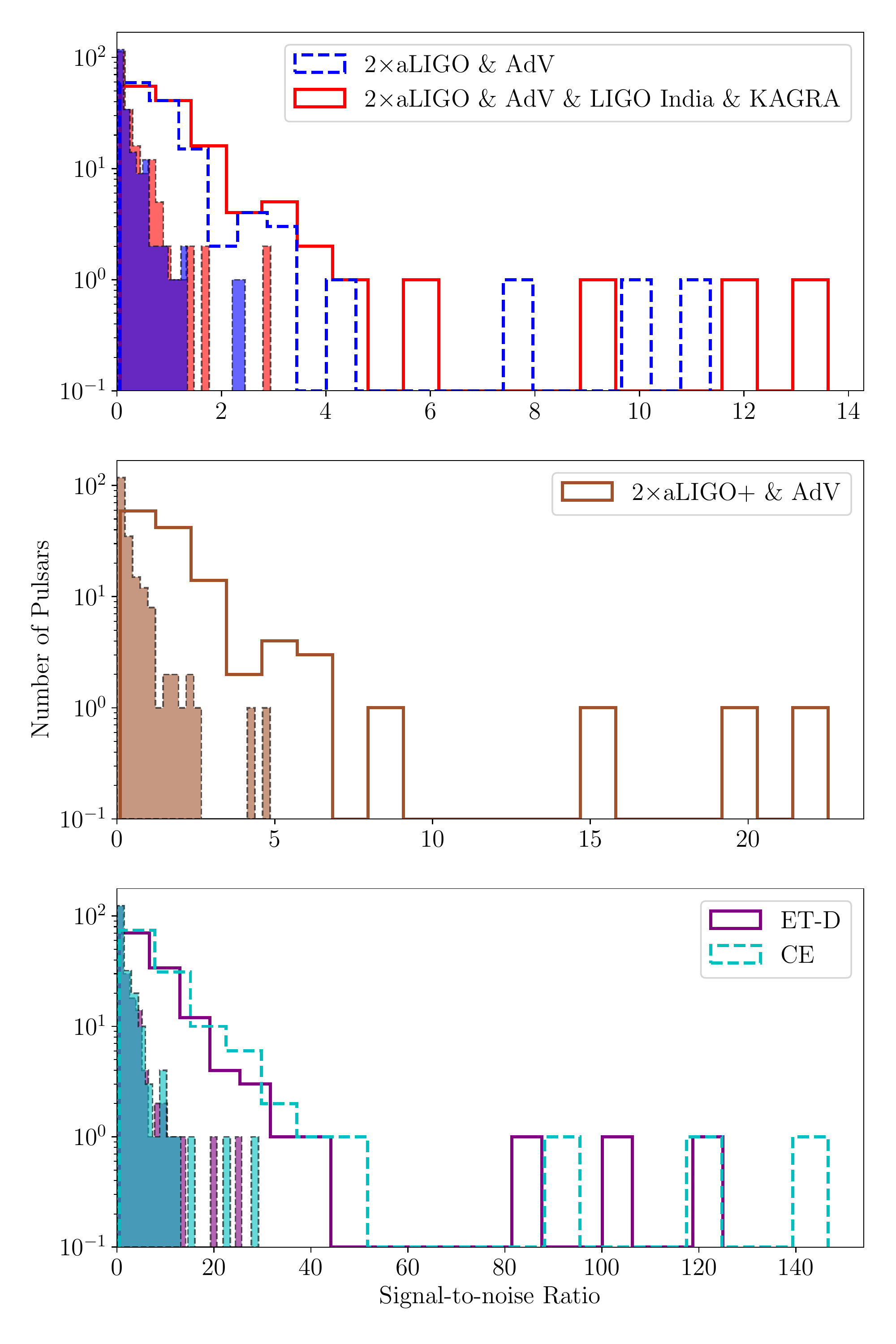}
\caption{\label{fig:snrs}TDistribution in signal-to-noise ratios of a
selection of MSPs, using a variety of future gravitational-wave observatory
networks with one year of coherent observations. The shaded histograms give the
distribution if we assume that all MSPs have ellipticities of $10^{-9}$ and the
canonical moment of inertia, and the unfilled histograms are the values if they
are all gravitars emitting at their spin-down limits.}
\end{figure}

The distributions of signal-to-noise ratios for all these scenarios are shown
in Figure~\ref{fig:snrs}, where the filled histograms represent sources all
having ellipticities of $10^{-9}$ and the canonical moment of inertia, and
the unfilled histograms show the distribution when the MSPs are gravitars. If
our minimum-ellipticity hypothesis is correct, we would expect the true
distribution to be somewhere between these two extremes, modulo the
uncertainties in distance and moment of inertia. The pulsar that would give
the highest signal-to-noise ratio with an ellipticity of $10^{-9}$ in all
these scenarios is PSR\,J1643\textminus1224, at a distance of 0.79\,kpc and
rotating at 216.4\,Hz. The pulsar that would have the highest signal-to-noise
ratio as a pure gravitar is PSR\,J0711\textminus6830, with a rotation
frequency of 182.1\,Hz and a distance of 0.11\,kpc. Figure~\ref{fig:snrs}
indicates that marginal detections from this MSP population could be possible
with the network of second-generation detectors, and more confident
detections possible with future third-generation detectors.

It is worth noting that, on the time scale of upgrades to aLIGO, the Square
Kilometre Array plans to start operation, bringing with it the possibility of
finding  $\sim 6\,000$ more MSPs \citep{2009A&A...493.1161S}. Pulsars that
occupy the current gap at the bottom-left of the P-Pdot diagram, i.e., with
short periods and very low spin-down rates, would provide strong evidence
against our hypothesis. Of course the hypothesis would be strongly supported if
the braking index of a boundary MSP was measured to be close to $n=5$. However
many decades or even centuries of observations, would most likely be required to
confidently determine the second period derivative necessary for this.

\acknowledgments
We thank the Institute for Nuclear Theory at the University of Washington, the
Department of Energy, and the organizers of the ``Astro-Solids, Dense Matter,
and Gravitational Waves'' workshop for their hospitality and for partial support
during completion of this work. M.P.\ is funded by the UK Science \& Technology
Facilities Council (STFC) under grant ST/N005422/1. P.D.L.\ is supported through
Australian Research Council (ARC) Future Fellowship FT160100112 and ARC
Discovery Project DP180103155. B.H.\ acknowledges support from the Polish National
Science Centre (SONATA BIS 2015/18/E/ST9/00577) and from the European Union's
Horizon 2020 research and innovation programme under grant agreement No.\ 702713.
D.I.J.\ acknowledges funding from STFC through grant No.\ ST/M000931/1. Partial
support comes from PHAROS, COST Action CA16214. The work has been made possible
using data from the ATNF Pulsar Catalog \citep{2005AJ....129.1993M}. This work
has been assigned LIGO document number LIGO-P1800125 and INT preprint number
INT-PUB-18-031.

\software{Jupyter \citep{kluyver2016jupyter}, NumPy \citep{numpy}, SciPy \citep{scipy}, Matplotlib \citep{Hunter:2007}, PSRQpy \citep{psrqpy}}

\bibliographystyle{aasjournal}
\bibliography{paper}

\end{document}